\documentclass[aps,reprint]{revtex4-2}
\usepackage{graphicx}
\usepackage{amsmath}
\usepackage{cases}
\usepackage{xcolor}

\draft % marks overfull lines with a black rule on the right

\begin{document}

%\preprint{}

\title{Deterministic time rewinding of waves in time-varying media}

\author{Seulong Kim}
\affiliation{Research Institute of Basic Sciences, Ajou University, Suwon 16499, Korea}

\author{Kihong Kim}
\email{khkim@ajou.ac.kr}
\affiliation{Department of Physics, Ajou University, Suwon 16499, Korea}
\affiliation{School of Physics, Korea Institute for Advanced Study, Seoul 02455, Korea}
\date{\today}
\begin{abstract}
Temporal modulation of material parameters offers unprecedented control over wave dynamics, enabling phenomena beyond the capabilities of static systems. Here we introduce and analyze a robust mechanism for time rewinding, whereby a temporally evolved wave is fully restored to its original state through a carefully engineered sequence of temporal modulations. In electromagnetic systems, time rewinding emerges from impedance-matched or anti-matched hierarchical bilayer structures with matched modulation durations, exploiting total transmission or reflection and reversed phase accumulation. In Dirac systems, it arises via complete interband transition driven by time-dependent vector potentials. Unlike time-reversal holography or quantum time mirrors, which produce wave echoes but only partial waveform recovery, our approach achieves deterministic and complete reconstruction of the entire wave state, including both amplitude and phase. Analytical conditions for robust amplitude and phase restoration are derived and validated through simulations of discrete and continuous modulations, demonstrating resilience to modulation complexity and temporal asymmetry. These findings establish a versatile platform for secure information retrieval, temporal cloaking, programmable metamaterials, and wave-based logic devices.
\end{abstract}

\maketitle

\section{Introduction}

Temporal modulation of material parameters provides a powerful means of controlling wave dynamics, enabling phenomena unattainable in static systems \cite{shvar,kal,nonreci,gali,most,asga}. Time-dependent variations in material properties give rise to scattering at temporal interfaces, generating forward- and backward-propagating spatial modes. With careful design, these processes facilitate precise manipulation of wave propagation, including deterministic time rewinding.

Recent advances in time-varying metamaterials have revealed a wide range of novel effects—such as momentum band gaps, temporal Anderson localization, nonreciprocal transport, and temporal analogs of Brewster phenomena \cite{zur,lust,tv8,tv9,tv10,akba,pac2,prud,tir2,lyu,solis,hors,kouts,riz,kouf,mir,kgap,shg1,shg2,pend1,tv2,miya,tv6,lustig,dong,jones,ren,shar,carm,apf,garn,jkim,eswa}. Building on this foundation, we investigate a distinct and highly tunable process: time rewinding, whereby a temporally evolved wave is fully restored to its original state through a tailored sequence of temporal modulations.

In electromagnetic systems, time rewinding arises from impedance-matched or anti-matched hierarchical bilayer structures with carefully matched modulation durations, leveraging total transmission or total reflection between time layers accompanied by reversed phase accumulation. In contrast, in Dirac systems where pseudospin dynamics and interband transitions are crucial \cite{dr1,dr2,sk1,ok,kim2}, time rewinding is achieved through complete interband transitions between matched temporal layers, driven by time-dependent vector potentials. Despite these differing physical mechanisms, both systems exhibit structurally analogous time-rewinding dynamics.

Notably, this process is fundamentally distinct from time-reversal holography \cite{tv10} and the quantum time mirror \cite{dr1}, which generate wave echoes through time reversal but only partially reconstruct the original waveform. In contrast, the mechanism described here enables deterministic and complete restoration of the entire wave state, including both amplitude and phase. The term {\it time rewinding} is inspired by the analogy of rewinding a video, in which previously recorded scenes are sequentially retrieved in reverse order.

To this end, we develop a general temporal scattering formalism for multilayer and continuous modulations, derive analytical criteria for full amplitude and phase recovery, and validate these predictions through numerical simulations of both discrete and continuous modulations, including full-field electromagnetic pulses. These results establish a robust and versatile platform for wave-based information retrieval, temporal cloaking, and programmable quantum and photonic devices.

\begin{figure}
\includegraphics[width=0.9\columnwidth]{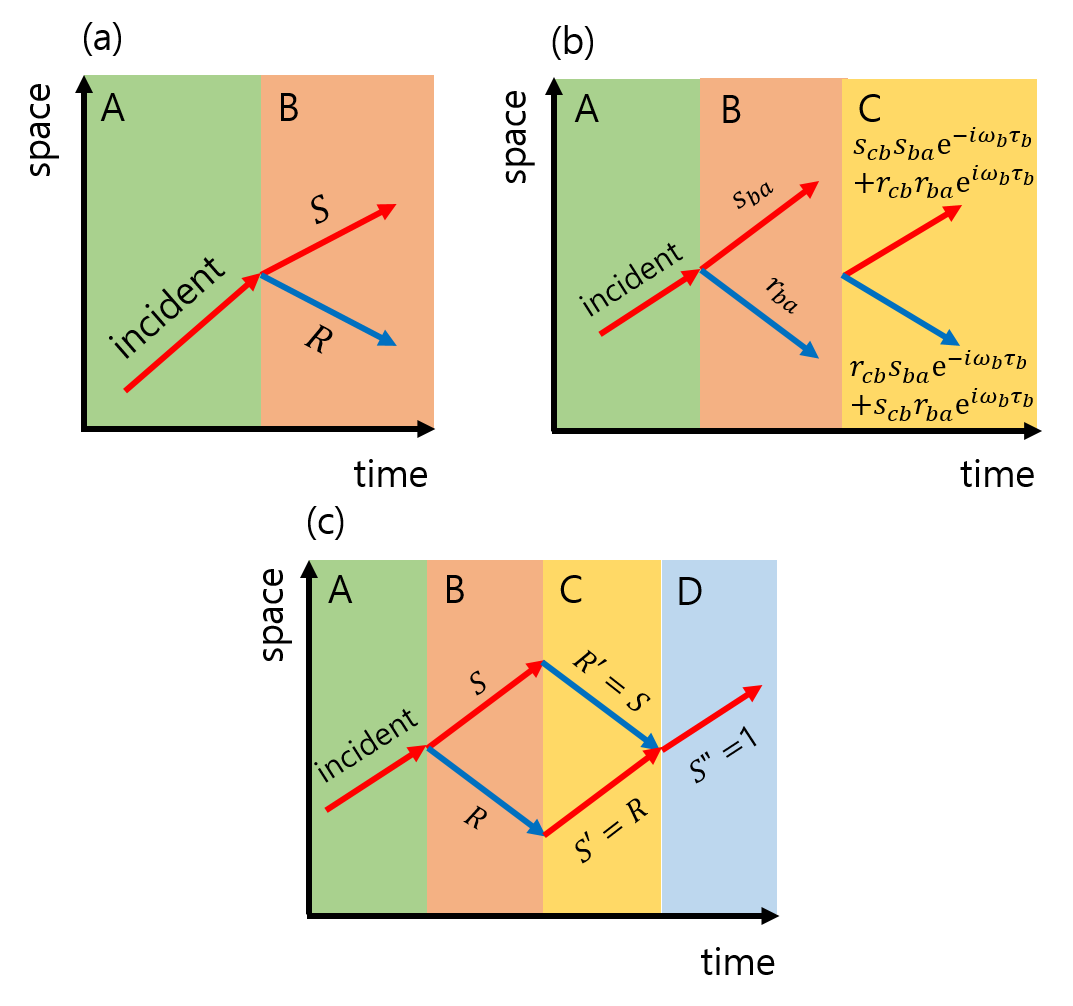}
\caption{ (a) Schematic of a temporal interface: a sudden change in material parameters induces temporal scattering, generating reflected and transmitted waves.
(b) Scattering at a temporal slab: total scattering is determined by interface coefficients and phase accumulation between interfaces.
(c) Temporal bilayer: impedance matching leads to total temporal transmission, while anti-matching results in total reflection. When applied sequentially with matched durations, scattering and phase effects cancel, restoring the initial wave state.}
\label{fig1}
\end{figure}

\section{Temporal scattering formalism}

We present a general scattering formalism for wave propagation in time-varying media, as schematically illustrated in Fig.~\ref{fig1}. In both electromagnetic and Dirac systems, temporal interfaces act analogously to spatial boundaries, inducing scattering into forward- and backward-propagating spatial modes. For multiple temporal interfaces—such as those in slabs, bilayers, or periodic configurations—the total scattering amplitudes are constructed from individual interface coefficients and the phase accumulations between them.

Consider two successive temporal interfaces at $t = t_1$ and $t = t_2 = t_1 + \tau$, dividing the time axis into three regions. The total forward ($s$) and backward ($r$) scattering amplitudes are given by
\begin{align}
s &= s_{21}s_{32}e^{-i\omega_2 \tau} + r_{21}r_{32}e^{i\omega_2 \tau}, \nonumber\\
r &= s_{21}r_{32}e^{-i\omega_2 \tau} + r_{21}s_{32}e^{i\omega_2 \tau},
\label{eq:bsf}
\end{align}
where $s_{ij}$ and $r_{ij}$ denote the transmission and reflection coefficients at the temporal interface from region $j$ to region $i$, and $\omega_2$ is the wave frequency in the intermediate region [see Fig.~\ref{fig1}(b)].
This formalism forms the basis for analyzing temporal multilayer structures and understanding phenomena such as time rewinding and temporal localization.

\subsection{Electromagnetic waves}

For electromagnetic waves propagating along the
$x$ axis in uniform isotropic media, the displacement field $D(t)$
satisfies
\begin{equation}
\frac{d}{dt}\left[\mu(t) \frac{d}{dt}D(t)\right] + \frac{c^2 k_x^2}{\epsilon(t)} D(t) = 0.
\label{eq:eme}
\end{equation}
Assuming harmonic solutions in each temporal region, the frequency is
given by $\omega_i = c k_x / n_i$, where  $n_i=\sqrt{\epsilon_i}\sqrt{\mu_i}$ is the refractive index, $k_x$ is the constant wave vector
along the $x$ axis, and $c$ is the speed of light in vacuum. The corresponding wave impedance is $\eta_i = \sqrt{\mu_i}/\sqrt{\epsilon_i}$.

At a temporal boundary from region 1 to 2, the scattering coefficients are
\begin{equation}
s_{21} = \frac{1}{2} \left(1 + \frac{\eta_1}{\eta_2}\right),~
r_{21} = \frac{1}{2} \left(1 - \frac{\eta_1}{\eta_2}\right).
\label{eq:emim}
\end{equation}
These expressions can be directly used in Eq.~(\ref{eq:bsf}), with the phase accumulation between interfaces determined by the corresponding frequency $\omega_i$ in each region.

\subsection{Dirac waves}

We consider the two-dimensional massless Dirac equation for a pseudospin-1/2 system with a two-component wavefunction
$\Psi=(\psi_1,\psi_2)^{\rm T}$,
subject to time-dependent scalar and vector potentials \cite{sk1}:
\begin{equation}
i\hbar \frac{d}{dt} \Psi(t) =
\begin{pmatrix}
U(t) &  \hbar v_Fq(t) \\
 \hbar v_Fq^*(t) & U(t)
\end{pmatrix} \Psi(t),
\label{eq:deq}
\end{equation}
where
\begin{equation}
q(t) = k_x + \frac{eA_x(t)}{\hbar} - i\left(k_y + \frac{eA_y(t)}{\hbar}\right)
\label{eq:qq}
\end{equation}
depends on the time-dependent vector potential
${\bf A}(t)$. Here, ${\bf{k}}=(k_x,k_y)$ is the conserved wave vector associated with the spatial part of the solution, and $v_F$ denotes the Fermi velocity.

At each temporal interface, scattering occurs between particle-like ($p$) and hole-like ($h$) states. The corresponding scattering coefficients $s_{ij}$ ($i,j=p,h$), representing transitions from state $j$ to $i$, are independent of the scalar potential $U(t)$ and given by
\begin{align}
&s_{pp} = s_{hh} = \frac{1}{2}\left(1 + \frac{f_2}{f_1}\right),\nonumber\\
&s_{ph} = s_{hp} = \frac{1}{2}\left(1 - \frac{f_2}{f_1}\right),
\label{eq:dim}
\end{align}
where $f_i = q_i / |q_i|$ in region $i$, with region 1 denoting the incident and region 2 the transmitted side \cite{sk1}. These coefficients can be inserted into the general formalism [Eq.~(\ref{eq:bsf})] to compute total scattering amplitudes across multilayer configurations, with phase accumulation governed by the frequency $\omega_i = v_F|q_i|$ in each region.

\section{Time-rewinding mechanism}

Time rewinding refers to the complete restoration of a wave that has undergone evolution due to temporal modulation, achieved through an engineered sequence of temporal medium transitions. In both electromagnetic and Dirac systems, this effect can be realized using temporal bilayers—pairs of conjugate temporal domains designed to cancel each other’s scattering and phase effects. This concept naturally extends to multilayer configurations, where a sequence of temporal layers is followed by its conjugate counterparts arranged in reverse order.

We consider a representative bilayer configuration, illustrated in Fig.~\ref{fig1}(c), where the temporal evolution of the medium follows the sequence
\begin{equation}
    A\xrightarrow{\text{at } t_1}B
    \xrightarrow{\tau_B} C \xrightarrow{\tau_C} D.\nonumber
\end{equation}
The system initially resides in medium $A$,
transitions to $B$ at time $t_1$, remains in $B$
for a duration $\tau_B$, then switches to $C$ for $\tau_C$,
and finally enters medium $D$.
The bilayer—comprising regions $B$ and $C$—is designed so that the scattering and phase effects induced by one layer are exactly canceled by those of the other, thereby acting as a temporal rewinder that restores the wave to its original state.

Although the specific conditions for time rewinding vary depending on the physical system, the underlying principle is universal: the second layer ($C$) must precisely reverse the dynamical effects introduced by the first layer ($B$). This evolution can be schematically summarized as
\begin{equation}
    \textrm {Wave~evolution:}~A\rightarrow B\rightarrow C\rightarrow D~\equiv~ A\rightarrow D.\nonumber
\end{equation}
In the following, we derive the explicit time-rewinding conditions for both electromagnetic and Dirac wave systems.

\subsection{Electromagnetic waves}

We now identify the specific conditions under which time rewinding can be achieved in electromagnetic systems. This effect can arise through two distinct mechanisms: total transmission enabled by impedance matching, and total reflection enabled by impedance anti-matching.

\vspace{1em}
\noindent \textbf{Case 1: Transmission-based time rewinding}

Transmission-based time rewinding occurs when total transmission is achieved at the interface between media $B$ and $C$, which requires matched impedances:
\begin{align}
\eta_B = \eta_C \quad \Rightarrow \quad s_{CB} = 1,~ r_{CB} = 0.
\end{align}
Additionally, the refractive indices must have opposite signs,
${\rm{sgn}}(n_B) = -{\rm{sgn}}(n_C)$,
and the durations must satisfy the condition
\begin{align}
\tau_B = \left\vert \frac{n_B}{n_C} \right\vert \tau_C.
\label{eq:dur}
\end{align}
When these conditions are met, the total phase accumulated in regions
$B$ and $C$ cancels exactly, restoring the wave to its initial state.
In isotropic media, this corresponds to a transition between a positive-index and a negative-index medium—that is, a case where both $\epsilon$ and $\mu$ are positive in one region and negative in the other. Notably, the magnitudes of $n_B$ and $n_C$ need not be equal; time rewinding is achieved by appropriately adjusting $\tau_C$ relative to $\tau_B$ according to Eq.~(\ref{eq:dur}).
We emphasize that this case represents a form of temporal perfect lensing, serving as the time-domain counterpart of spatial perfect lensing realized using negative-index media \cite{pendry}.

It is important to note that transmission-based time rewinding is not equivalent to temporal matching phenomena studied in \cite{pac2,rama1,rama2}, which refer to the absence of scattering at a temporal interface due to impedance matching. In our case, while impedance matching is necessary, it is not sufficient. Time rewinding additionally requires a sign change in the refractive index and must satisfy the duration constraint of Eq.~\eqref{eq:dur}. These stricter conditions enable active reversal of the wave’s temporal evolution, distinguishing our mechanism from standard temporal matching.

\vspace{1em}
\noindent \textbf{Case 2: Reflection-based time rewinding}

Reflection-based time rewinding relies on total reflection at the $B \rightarrow C$ interface, which occurs when the impedances have equal magnitudes but opposite signs:
\begin{align}
\eta_B = -\eta_C \quad \Rightarrow \quad s_{CB} = 0, \quad r_{CB} = 1.
\end{align}
In addition, the refractive indices must have the same sign.
As in the transmission-based case, the duration condition given by Eq.~\eqref{eq:dur} must also be satisfied to achieve complete time rewinding. Although the wave is reflected rather than transmitted, the accumulated phase is exactly reversed, resulting in complete restoration of the initial wave state.
This case corresponds to media in which $\epsilon$ is positive and $\mu$ is negative in region $B$, and $\epsilon$ is negative and $\mu$ is positive in region $C$, or vice versa. In such cases, both the impedance and refractive index become purely imaginary: the impedances have opposite signs, while the refractive indices share the same sign.
A spatial analog of this configuration has previously been studied in the context of tunneling through conjugate-matched pairs \cite{alu00}.

In both mechanisms, once the time-rewinding conditions are satisfied, the total scattering coefficients for the four-region structure reduce to those of a direct transition from medium $A$ to medium $D$:
\begin{align}
s_{\mathrm{total}} = \frac{1}{2} \left( 1 + \frac{\eta_A}{\eta_D} \right), ~
r_{\mathrm{total}} = \frac{1}{2} \left( 1 - \frac{\eta_A}{\eta_D} \right).
\end{align}
Thus, the presence of the bilayer $B$ and $C$ has no effect on the scattering coefficients.
These two mechanisms—impedance matching and anti-matching—serve as temporal analogs of spatial wave phenomena such as omnidirectional surface wave excitation and super-Klein tunneling between two distinct bi-isotropic media \cite{Kim20}.

\subsection{Dirac waves}

It is instructive to explore formal analogies between the electromagnetic and Dirac cases.
The pseudospin-1/2 Dirac equation, Eq.~(\ref{eq:deq}), consists of two coupled first-order differential equations in time. By eliminating $\psi_2$, we obtain a second-order equation for $\psi_1$:
\begin{align}
\frac{d}{dt}\left[\frac{1}{q(t)}\frac{d}{dt}\psi_1(t)\right]+v_F^2q^*(t)\psi_1(t)=0.
\end{align}
Comparing this with the electromagnetic wave equation, Eq.~(\ref{eq:eme}), we identify the following correspondences:
\begin{align}
    \psi_1(t)\leftrightarrow D(t),~\frac{1}{q(t)} \leftrightarrow \mu(t),~v_F^2q^*(t)\leftrightarrow\frac{c^2 k_x^2}{\epsilon(t)}.
\end{align}
These analogies lead to expressions for the effective impedance and refractive index in the Dirac case:
\begin{align}
    \eta=\frac{\sqrt{\mu}}{\sqrt{\epsilon}}\leftrightarrow \sqrt{\frac{q^*}{q}},~n=\sqrt{\epsilon}\sqrt{\mu}\leftrightarrow \frac{1}{\vert q\vert}.
\end{align}
Within this framework, total transmission under impedance matching and total reflection under impedance anti-matching correspond to complete intraband and interband transitions, respectively, in the Dirac system \cite{sk1}.
However, a crucial distinction arises: the effective refractive index in the Dirac case, given by $1/|q|$, is always a positive real number and cannot change sign. This prohibits time rewinding via transmission, which in electromagnetic systems relies on both impedance matching and sign reversal of the refractive index.
As a result, while perfect transmission is theoretically possible in the Dirac case through impedance matching, it does not constitute time rewinding due to the absence of index inversion. In contrast, perfect time rewinding remains achievable in both electromagnetic and Dirac systems via total reflection enabled by impedance anti-matching.

We now examine the specific conditions under which temporal scattering in Dirac systems can be reversed. A key requirement is complete interband conversion, which occurs when the direction of the complex wave vector
$q(t)$ is inverted, i.e., when $f_C/f_B = -1$.
Under this condition, the interband transition coefficients between $B$ and $C$, $s_{ph}$ and $s_{hp}$, become unity, while the intraband terms $s_{pp}$ and $s_{hh}$ vanish.
As a result, a $p$-band state is fully converted into a $h$-band state, and vice versa.

In the absence of a scalar potential, time rewinding is achieved when the durations satisfy
\begin{align}
\tau_B = \left\vert \frac{q_C}{q_B} \right\vert \tau_C,
\label{eq:ddur}
\end{align}
ensuring that the phases accumulated in regions
$B$ and $C$ cancel exactly. This condition is directly analogous to the impedance anti-matching condition in electromagnetic systems, as seen by comparing Eqs.~(\ref{eq:emim}) and (\ref{eq:dim}).

When a scalar potential $U(t)$ is present, the effective scattering coefficients for the full four-region sequence are
\begin{align}
&s_{pp}^{\mathrm{total}} =s_{hh}^{\mathrm{total}} =\frac{1}{2}\left(1 + \frac{f_D}{f_A}\right) e^{-i (U_B \tau_B + U_C \tau_C)/\hbar}, \nonumber\\
&s_{ph}^{\mathrm{total}} = s_{hp}^{\mathrm{total}} =\frac{1}{2}\left(1 - \frac{f_D}{f_A}\right) e^{-i (U_B \tau_B + U_C \tau_C)/\hbar}.
\end{align}
If $U_B=-U_C$, these expressions reduce to those of a single interface between regions $A$ and $D$, indicating full restoration of the wavefunction in both amplitude and phase. Even when the scalar potential varies arbitrarily, only the global phase is affected, and the probability density remains unchanged.

\begin{figure}
\includegraphics[width=0.9\columnwidth]{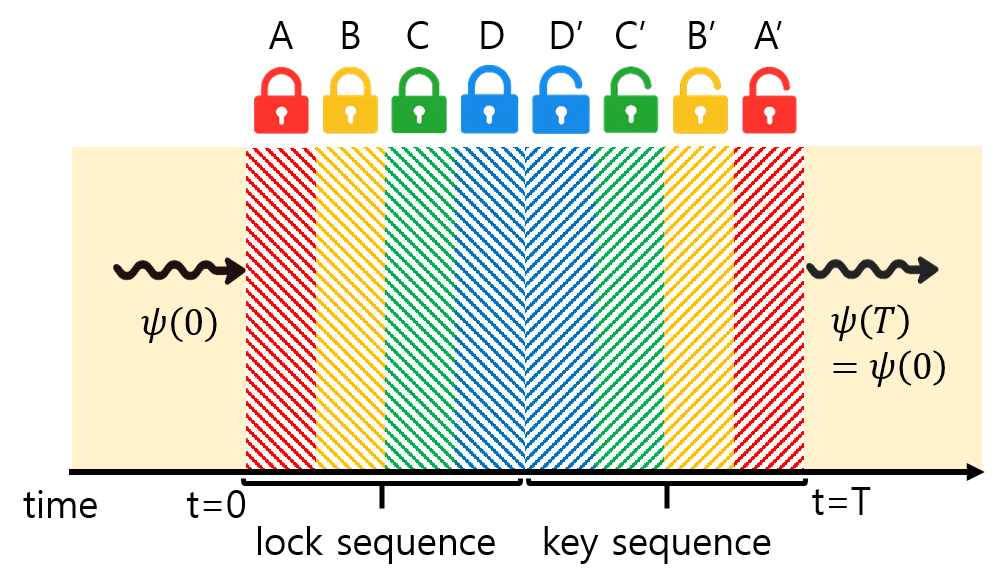}
\caption{Schematic illustration of time rewinding in a temporal multilayer system.
Each conjugate pair—$(A, A')$, $(B, B')$, $(C, C')$, and $(D, D')$—is designed to satisfy the time-rewinding matching conditions.}
\label{fig2}
\end{figure}

\subsection{Time rewinding in multilayers}

Regardless of whether the underlying mechanism involves total transmission, total reflection, or complete interband transition, the core principle remains the same: the second temporal medium in a bilayer is engineered to precisely cancel the scattering and phase effects introduced by the first, resulting in full restoration of the wave state, including both amplitude and phase. This approach is analytically tractable, robust against perturbations, and naturally extends to multilayer temporal configurations.

The cancellation mechanisms described above are not limited to individual bilayers but can be systematically extended to complex multilayer configurations. When each temporal modulation is followed by its conjugate counterpart—with matched or anti-matched impedances in electromagnetic systems, or conjugate wave vectors $q$ in Dirac systems, and appropriately scaled durations—the entire structure functions as an effective time-rewinding operator. For instance, consider the sequence illustrated in Fig.~\ref{fig2}:
\begin{equation}
    A \to B \to C \to D \to D^\prime \to C^\prime \to B^\prime \to A^\prime.
\end{equation}
If each conjugate pair $(A,A^\prime)$, $(B,B^\prime)$, $(C,C^\prime)$, and $(D,D^\prime)$ satisfies the corresponding time-rewinding matching conditions, the cumulative effects of all intermediate layers cancel. The overall evolution is then equivalent to a direct transition from the initial to final state, enabling complete recovery of the original wave state even in complex temporally modulated systems.

We emphasize that the media before $t=0$ and after $t=T$ need not be identical. The ability to time-rewind does not rely on identical input/output layers; it is achieved when the intermediate temporal multilayer implements the inverse temporal evolution of the forward process. In a spatially uniform, time-varying medium the temporal layers exchange energy with the fields, so the instantaneous energies in two conjugate layers within the rewinding segment (say
$A$ and $A^\prime$) need not be equal, even if the fields are the same.
What the rewinding segment guarantees is that the field state just before the final jump equals the time-reversed field at the start of the segment. The energy at these two instants can differ because it is evaluated with the material parameters of the layer. If the initial and final layers are identical and rewinding is exact, then after the last jump the energy equals the initial energy; just before that jump, no such equality is implied.

Although the analysis above assumes abrupt temporal transitions, the formalism naturally extends to media with smoothly varying temporal profiles. In such cases, the same time-rewinding conditions apply, provided that the modulation sequence and its conjugate counterpart satisfy the appropriate cancellation criteria. Explicit calculations for smoothly varying media can be performed using the invariant imbedding method, as detailed in Appendix for both electromagnetic and Dirac systems.

\begin{figure}
  \centering
  \includegraphics[width=0.9\columnwidth]{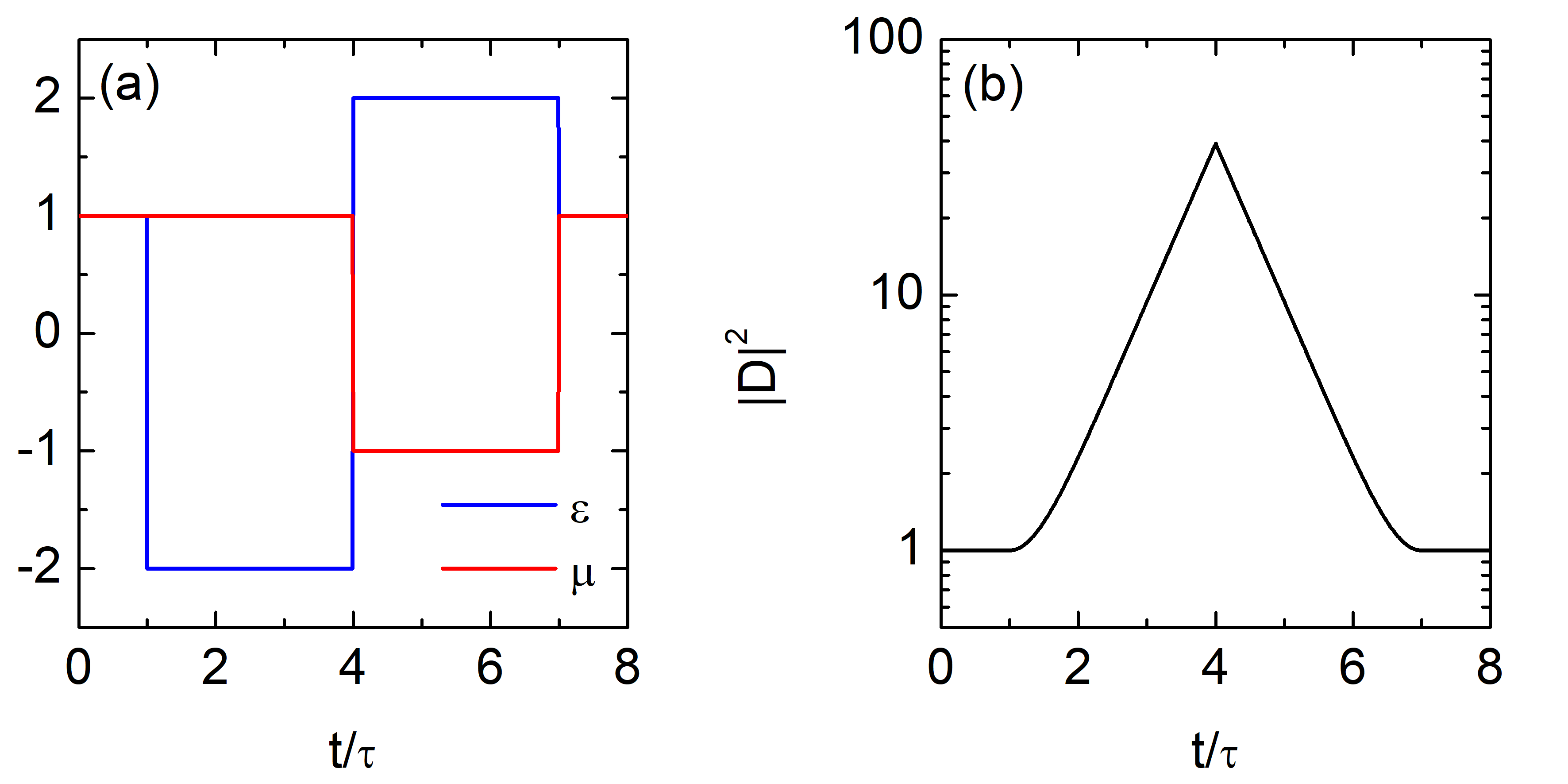}
  \caption{(a) Temporal profiles of permittivity $\epsilon(t)$ and permeability $\mu(t)$ in a simple bilayer time-varying medium, with $\tau = (ck_x)^{-1}$.
(b) Formation of a temporally localized wave: the wave intensity grows exponentially in the first slab and decays in the second, due to identical refractive indices but anti-matched impedances, producing a localized peak at $t = 4\tau$.}
\label{fig3}
\end{figure}

\begin{figure}
  \centering
  \includegraphics[width=0.9\columnwidth]{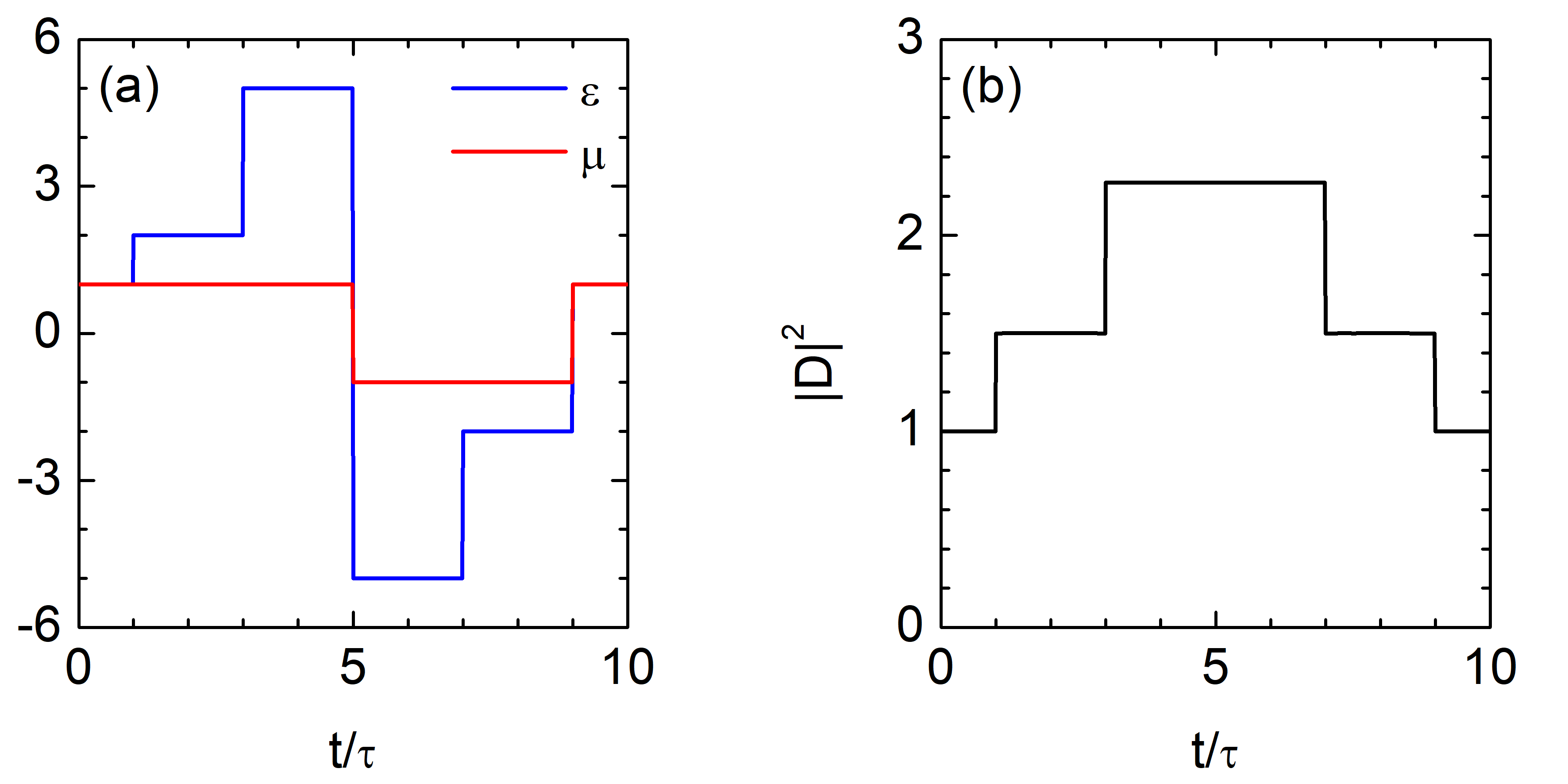}
  \caption{(a) Temporal profiles of permittivity $\epsilon(t)$ and permeability $\mu(t)$ in a four-layer time-varying medium, with $\tau = (ck_x)^{-1}$.
(b) The electric displacement field, amplified by temporal scattering, is fully restored at $t = 9\tau$ through the time-rewinding mechanism.}
\label{fig4}
\end{figure}

\begin{figure}
  \centering
  \includegraphics[width=0.9\columnwidth]{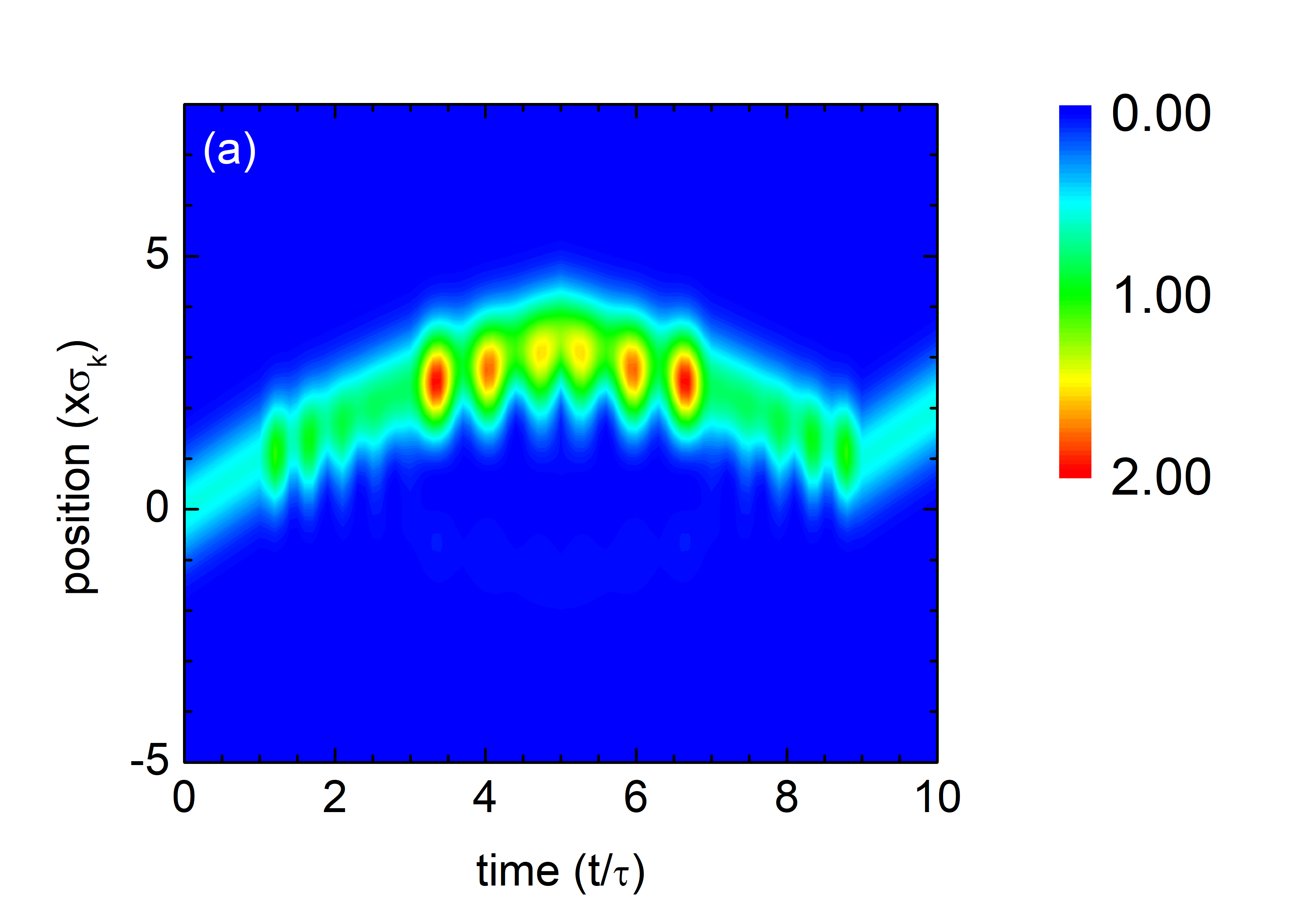}
  \includegraphics[width=0.9\columnwidth]{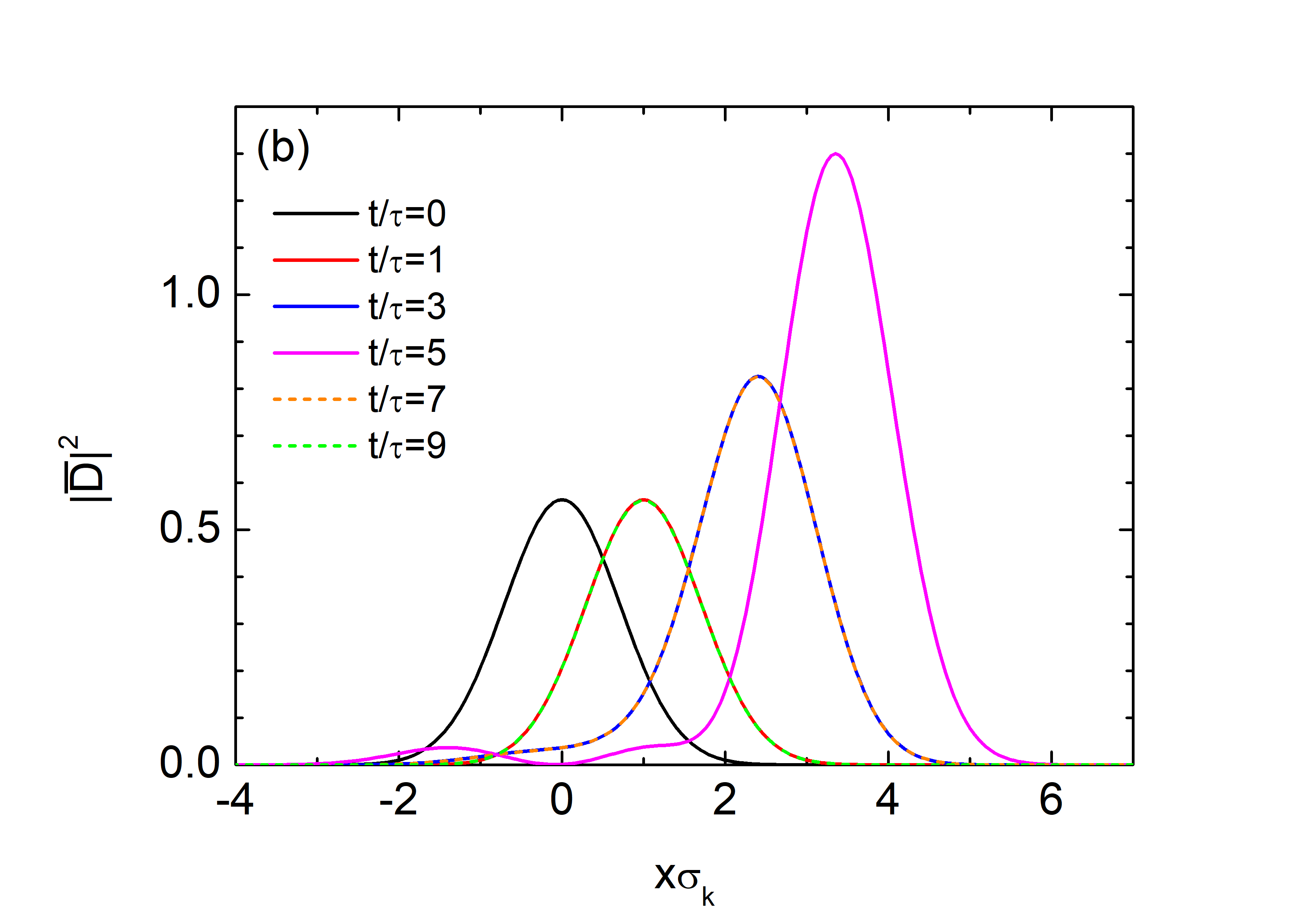}
  \caption{
(a) Contour plot of a Gaussian pulse with spectral width $\sigma_k$ and central wavenumber $k_c = 10\sigma_k$, shown as a function of space and time while propagating through the temporal structure in Fig.~\ref{fig4}(a). The pulse retraces its trajectory, demonstrating complete waveform restoration via time rewinding. The characteristic time scale is defined as $\tau = (c\sigma_k)^{-1}$.
(b) Spatial profiles of the pulse at selected times. The normalized displacement field intensity is given by $|\bar{D}|^2 = |D|^2 / \int_{-\infty}^{\infty} |u(x,0)|^2 dx$.
}
\label{fig5}
\end{figure}

\section{Numerical results}

We validate the proposed time-rewinding mechanisms through numerical simulations of wave propagation in both electromagnetic and Dirac systems. The results demonstrate complete recovery of the initial wave state—even in multilayer configurations, under continuous temporal modulation, and in the presence of additional perturbations. All simulations were performed using a Fortran code developed by the authors, based on a generalized invariant imbedding method capable of handling arbitrary temporal variations.

\subsection{Time rewinding of electromagnetic waves}

We investigate time rewinding in electromagnetic systems using the theoretical framework developed above. Simulations are performed in a multilayer medium with temporally modulated permittivity and permeability, focusing on the reversal of a propagating pulse in a dispersionless regime. Complete time rewinding is verified through the full recovery of both wave amplitude and phase.

As an initial demonstration, we show that wave evolution can be reversed using a simple temporal bilayer satisfying the impedance anti-matching condition. As noted in \cite{Pach}, wave propagation is suppressed and the amplitude grows exponentially when the refractive index becomes imaginary. Leveraging this effect, we consider the following temporal profiles:
\begin{align}
\epsilon(t) &=
\begin{cases}
-2, & \tau < t \le 4\tau \\
2, & 4\tau < t \le 7\tau \\
1, & \text{otherwise}
\end{cases} ,\nonumber\\
\mu(t) &=
\begin{cases}
-1, & 4\tau < t \le 7\tau \\
1, & \text{otherwise}
\end{cases}.
\end{align}
These profiles, shown in Fig.~\ref{fig3}(a), produce imaginary refractive indices and impedances during both intervals $\tau < t \le 4\tau$ and $4\tau < t \le 7\tau$, with the impedances equal in magnitude but opposite in sign—i.e., anti-matched. As a result, the wave undergoes exponential amplification in the first slab and symmetric exponential decay in the second. This creates a temporally localized wave centered at $t = 4\tau$, characterized by an exponential temporal envelope, as shown in Fig.~\ref{fig3}(b). Full recovery of both amplitude and phase confirms successful time rewinding. The emergence of a temporal surface wave during the intermediate stage, a large-amplitude waveform confined to a narrow time window, is a distinctive feature of the reflection-based time-rewinding mechanism.

As noted earlier, the time-rewinding effect is not limited to a single bilayer system but can be generalized to multilayered structures, provided each constituent bilayer satisfies the necessary conditions for time rewinding. To demonstrate the generality of this mechanism, we consider a four-layer temporal structure designed to satisfy two distinct impedance-matching time-rewinding conditions, as illustrated in Fig.~\ref{fig4}(a). The time intervals $\tau < t \le 3\tau$ and $7\tau < t \le 9\tau$, as well as $3\tau < t \le 5\tau$ and $5\tau < t \le 7\tau$, each form a matched bilayer: they exhibit equal impedances, opposite refractive indices, and equal durations. As a result, the amplified electric displacement field generated by temporal scattering is fully restored at $t = 9\tau$ via the time-rewinding mechanism, as shown in Fig.~\ref{fig4}(b).

In the absence of dispersion, a pulse can be fully reversed, with all momentum components recovering their original amplitude and phase. Figure~\ref{fig5} shows the evolution of a Gaussian pulse propagating through the temporal multilayer structure of Fig.~\ref{fig4}. The initial pulse is given by
\begin{equation}
u(x,0) = \int_{-\infty}^{\infty} \exp\left[ -\frac{(k_x - k_c)^2}{2\sigma_k^2} + i k_x x \right] dk_x,
\end{equation}
where $\sigma_k$ is the spectral width and $k_c = 10\sigma_k$ is the central wavenumber. The parameters are chosen such that $c\sigma_k\tau = 1$.

For $t < 5\tau$, the pulse travels in the $+x$ direction. After $t = 5\tau$, negative refraction in the subsequent time layers reverses the group velocity, redirecting the pulse toward $-x$. Although the medium is spatially homogeneous and momentum is conserved, both group and phase velocities reverse.
Figure~\ref{fig5}(a) presents a space-time contour plot of the pulse, showing that it retraces its trajectory and is fully restored at $t = 9\tau$, i.e., $u(x,9\tau) = u(x,\tau)$. The normalized displacement field intensity is defined as $|\bar{D}(t)|^2 = |D(t)|^2 / \int_{-\infty}^{\infty} |u(x,0)|^2 dx$. Figure~\ref{fig5}(b) confirms time rewinding, with matching profiles at $t = \tau$ and $9\tau$, and at $3\tau$ and $7\tau$, demonstrating complete waveform recovery.

\begin{figure}
\includegraphics[width=0.9\columnwidth]{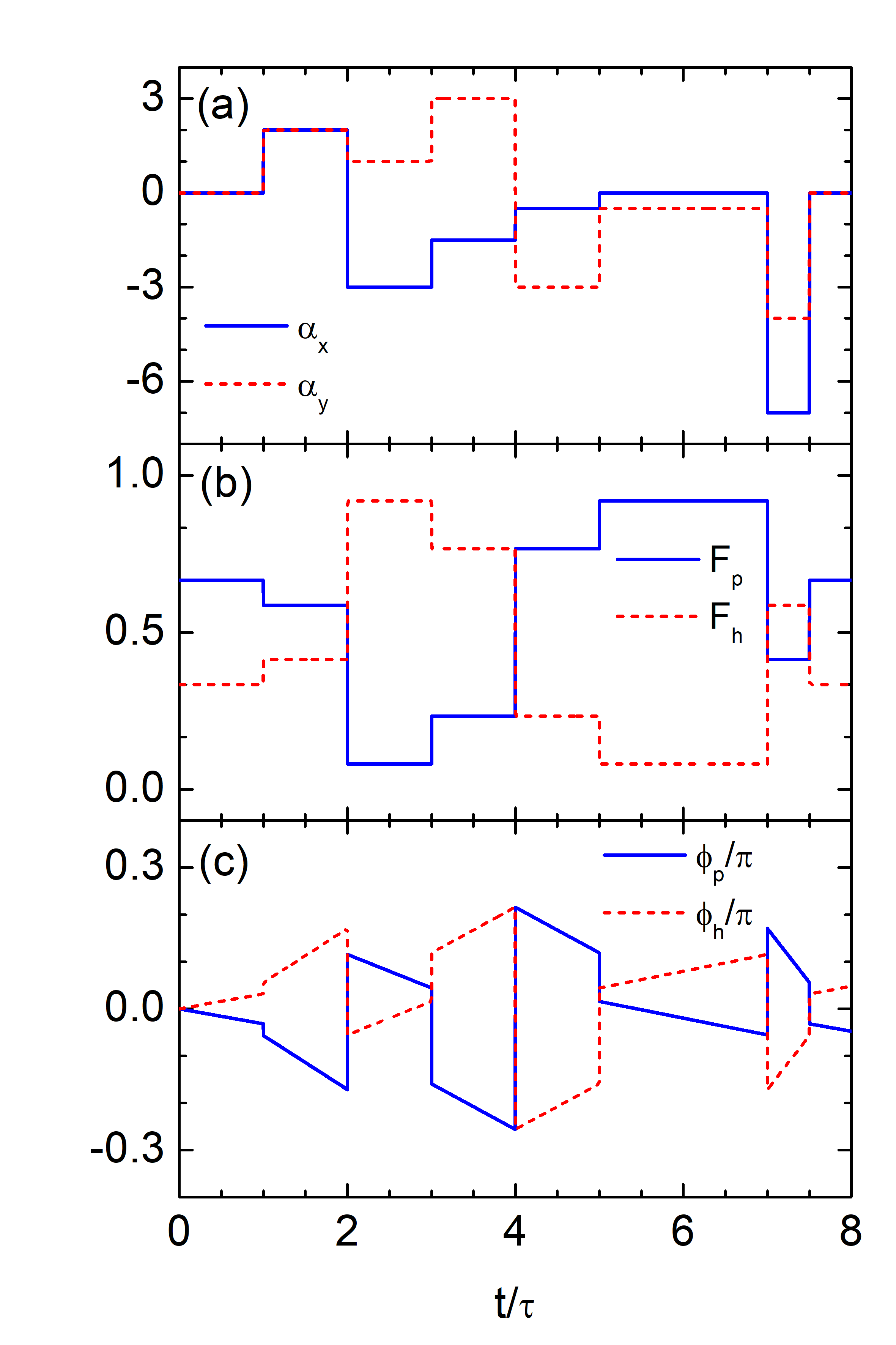}
\caption{Time rewinding of Dirac waves in a multilayer temporal structure.
(a) Temporal profiles of the normalized vector potential components $\alpha_x$ and $\alpha_y$ from $t = \tau$ to $t = 7.5\tau$ across six layers with durations $\tau$, $\tau$, $\tau$, $\tau$, $2\tau$, and $0.5\tau$, where $\tau = 0.1/(kv_F)$.
(b, c) Probability densities ($F_p$, $F_h$) and phases ($\phi_p$, $\phi_h$) of the wavefunction in each band. The modulations in the first three layers are precisely compensated by those in the last three, resulting in complete recovery of the initial wave state.
}
\label{figd1}
\end{figure}

\begin{figure}
\includegraphics[width=0.9\columnwidth]{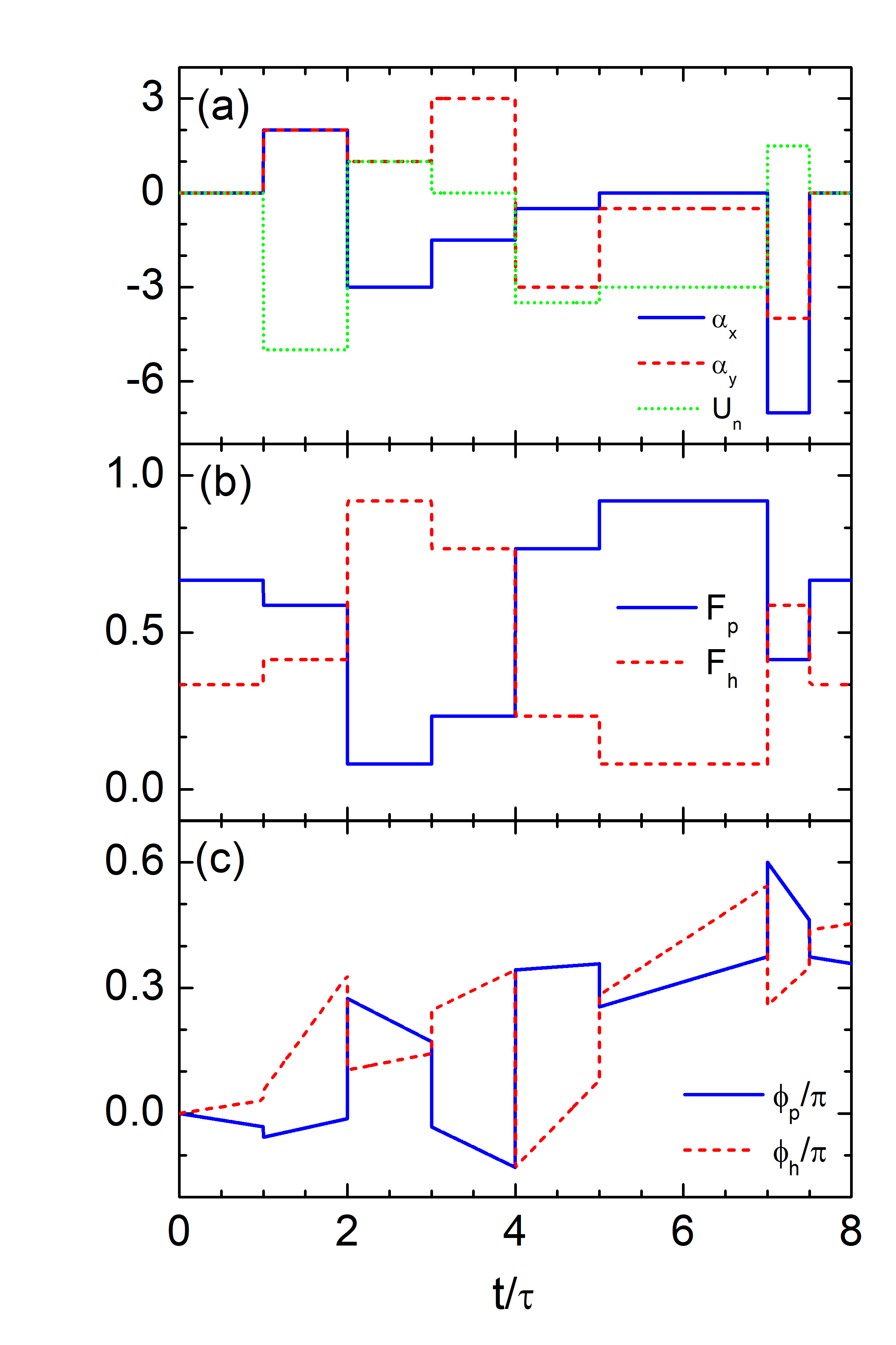}
\caption{Time rewinding of Dirac waves across multiple temporal layers with time-dependent scalar and vector potentials.
(a) Temporal profiles of the normalized scalar ($U_n$) and vector ($\alpha_x$, $\alpha_y$) potentials over six layers from $t = \tau$ to $t = 7.5\tau$.
(b) Probability density of the wavefunction in each band, showing that the effects of the first three modulations are fully compensated by the subsequent three, despite the presence of the scalar potential.
(c) Phase of the wavefunction, which fails to recover due to the additional scalar potential, indicating a loss of phase coherence.
}
\label{figd2}
\end{figure}

\begin{figure}
\includegraphics[width=0.9\columnwidth]{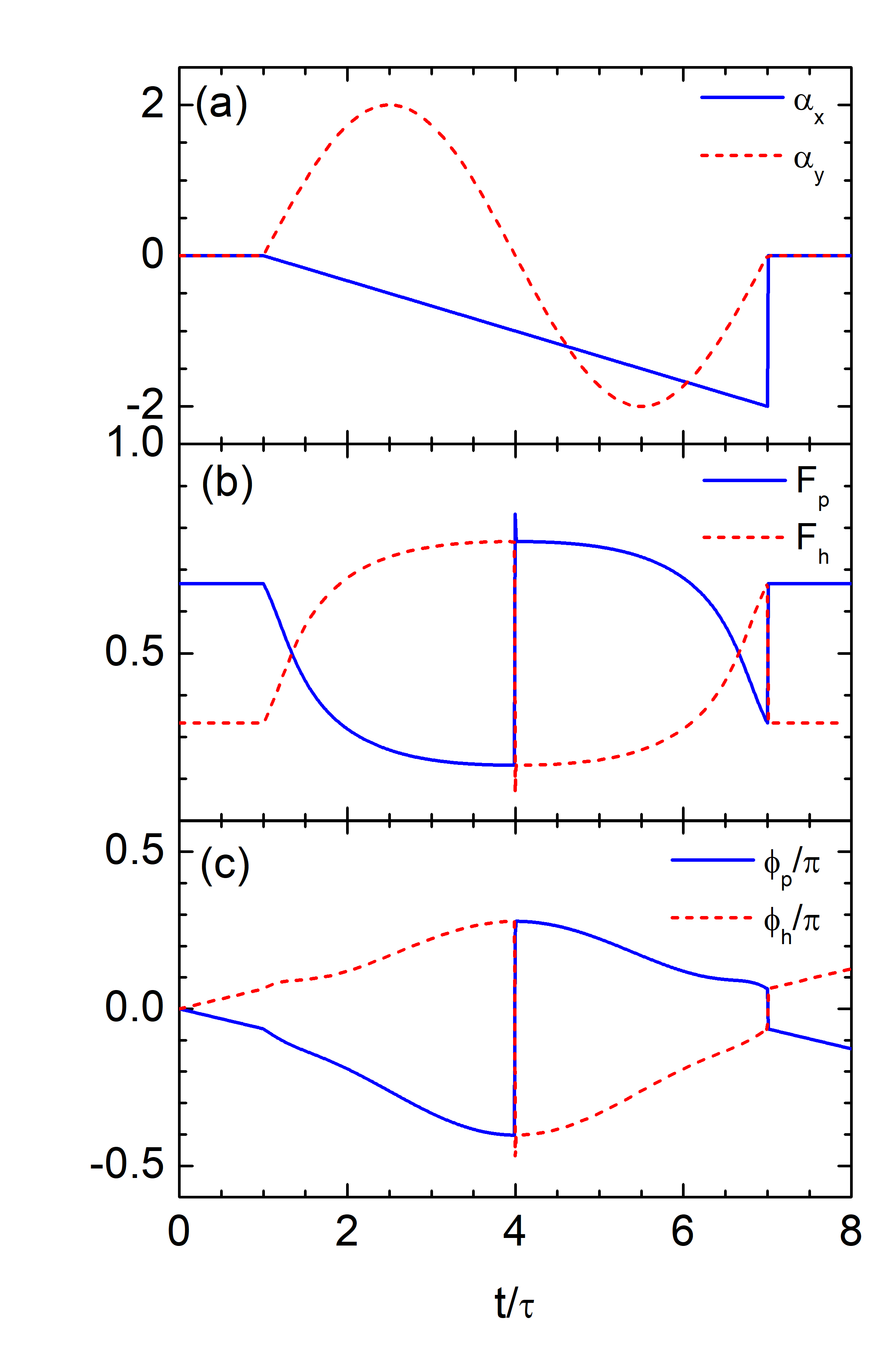}
\caption{Time rewinding under continuous temporal modulation of the vector potential.
(a) Temporal profile of the vector potential, which evolves smoothly from $t = \tau$ and returns to zero at $t = 7\tau$.
(b) Probability density of the wavefunction in each band, showing full recovery of the initial values at $t = 7\tau$.
(c) Phase of the wavefunction, also restored to its initial value, confirming that both amplitude and phase are fully recovered through the continuous modulation.}
\label{figd3}
\end{figure}

\subsection{Time rewinding of Dirac waves}

We now turn to Dirac waves.
Figure~\ref{figd1} illustrates time rewinding achieved through a temporal multilayer sequence. In Fig.~\ref{figd1}(a), the normalized vector potential components $\alpha_x = eA_x/(\hbar k)$ and $\alpha_y = eA_y/(\hbar k)$, with $k$ denoting the magnitude of the wave vector, are modulated from $t = \tau$ to $t = 7.5\tau$. The sequence comprises six layers with nonuniform durations: $\tau$, $\tau$, $\tau$, $\tau$, $2\tau$, and $0.5\tau$, where $\tau = 0.1/(kv_F)$. These results demonstrate that symmetric layer durations are not necessary for time rewinding, provided the matching conditions are satisfied. Each layer’s duration can be adjusted relative to the vector potential magnitude, or vice versa, according to Eq.~(\ref{eq:ddur}).
Figures~\ref{figd1}(b) and \ref{figd1}(c) show the resulting probability densities
\begin{align}
F_p = |s_{pp}|^2, \quad F_h = |s_{hp}|^2,
\end{align}
and phase angles $\phi_p$ and $\phi_h$ associated with $s_{pp}$ and $s_{hp}$, respectively. These results confirm that the temporal evolution induced by the first three layers are exactly compensated by the last three, leading to complete restoration of the initial wave state. At the final time $t = 7.5\tau$, the system satisfies $F_p(7.5\tau) = F_p(\tau)$, $F_h(7.5\tau) = F_h(\tau)$, $\phi_p(7.5\tau) = \phi_p(\tau)$, and $\phi_h(7.5\tau) = \phi_h(\tau)$, indicating perfect recovery of both amplitude and phase.

To evaluate the robustness of the time-rewinding mechanism, we introduce a normalized time-dependent scalar potential $U_n = U/(\hbar kv_F)$ while keeping the vector potential profile unchanged, as shown in Fig.~\ref{figd2}(a). Despite this perturbation, the probability densities remain fully restored [Fig.~\ref{figd2}(b)], confirming that complete interband transitions continue to enable amplitude recovery. However, phase coherence is no longer preserved: Fig.~\ref{figd2}(c) shows that $\phi_p(7.5\tau) \ne \phi_p(\tau)$ and $\phi_h(7.5\tau) \ne \phi_h(\tau)$, indicating that the scalar potential induces a global phase shift. Thus, while amplitude recovery is robust to scalar potential perturbations, the phase is sensitive, and coherence is lost in the more general case.

As discussed earlier, time rewinding is not limited to systems with discrete temporal discontinuities—it can also occur in media with continuously varying properties. To illustrate this, we consider a smoothly time-dependent vector potential defined as
\begin{align}
&\alpha_x(t) = \begin{cases}
                    \frac{1}{3}\left(1-\frac{t}{\tau}\right), & \mbox{if } \tau<t\le 7\tau \\
                    0, & \mbox{otherwise}
                  \end{cases}, \\
&\alpha_y(t) = \begin{cases}
                    2\sin\left[\frac{\pi}{3}(\frac{t}{\tau}-1)\right], & \mbox{if } \tau<t\le 7\tau \\
                    0, & \mbox{otherwise}
                  \end{cases}.
\end{align}
As shown in Fig.~\ref{figd3}, the vector potential starts at zero, varies smoothly from $t = \tau$, and returns to zero at $t = 7\tau$. This continuous modulation drives a gradual evolution of the band populations $F_p$ and $F_h$. A complete interband transition occurs at $t = 4\tau$, after which the system retraces its prior evolution. At the final time $t = 7\tau$, both amplitude and phase are fully restored, satisfying $F_p(7\tau) = F_p(\tau)$, $F_h(7\tau) = F_h(\tau)$, $\phi_p(7\tau) = \phi_p(\tau)$, and $\phi_h(7\tau) = \phi_h(\tau)$. These results confirm that time rewinding can be achieved even under continuous temporal modulation.

\section{Discussion}

We have developed a unified framework for deterministic time rewinding in electromagnetic and Dirac systems using temporally engineered structures. In both cases, carefully designed modulations cancel accumulated scattering and phase, enabling complete recovery of the initial wave state, including both amplitude and phase. This effect extends beyond discrete transitions to smoothly varying modulation profiles, highlighting the robustness and generality of the mechanism.

In photonic systems, time rewinding is achieved through impedance-matched or anti-matched temporal layers. We demonstrate that temporally localized surface waves can emerge in media with imaginary refractive indices, establishing a temporal analog of spatial boundary phenomena. Broadband pulses in dispersionless regimes can also be fully rewound, allowing for coherent temporal shaping and waveform correction.

This mechanism enables applications in secure communications, temporal cloaking, programmable nanophotonic circuits, and wave-based logic devices. Its analytical transparency and compatibility with dynamic media position it as a promising approach for reconfigurable, next-generation photonic and quantum platforms.

Looking ahead, integrating temporal modulation with nonlinear, dissipative, and non-Hermitian effects could unlock entirely new classes of adaptive photonic devices.

\begin{acknowledgments}
This research was supported by the Basic Science Research Program through the
National Research Foundation of Korea (https://ror.org/013aysd81) funded by the Ministry of Education
(RS-2021-NR060141).
\end{acknowledgments}

\appendix

\section*{Appendix: Invariant imbedding method for electromagnetic and Dirac waves in time-varying media}
\renewcommand{\theequation}{A\arabic{equation}}
\setcounter{equation}{0}

In this appendix, we present the invariant imbedding method as a framework for analyzing wave scattering induced by continuous temporal variations in the medium. While widely applied to spatially varying systems \cite{sk2,sk3}, we extend its formulation here to accommodate time-dependent media \cite{kim2}.

\subsection*{Electromagnetic waves}

We begin with the case of electromagnetic waves. In a stationary medium, the field components evolve as $\exp[i(k_xx \mp \omega t)]$. When the medium parameters vary in time, scattering occurs at temporal interfaces, generating reflected and transmitted components. Consider a unit-amplitude plane wave incident from the past ($t < 0$) in the $+x$ direction. If the permittivity $\epsilon(t)$ and permeability $\mu(t)$ vary within the interval $0 \le t \le T$, the electric displacement field takes the form
\begin{equation}
D(t) =
\begin{cases}
e^{-i\omega_1 t}, & t < 0 \\
s(T) e^{-i\omega_2 (t - T)} + r(T) e^{i\omega_2 (t - T)}, & t > T
\end{cases},
\end{equation}
where $\omega_1 = c k_x / \sqrt{\epsilon_1\mu_1}$ and $\omega_2 = c k_x / \sqrt{\epsilon_2\mu_2}$, with $\epsilon_{1,2}$ and $\mu_{1,2}$ denoting the initial and final material parameters, respectively.

To derive the invariant imbedding equations, we introduce an auxiliary field $u_1(t) = D(t)/s(T)$, leading to
\begin{equation}
 u_1(t)=
 \begin{cases}
   \rho_i(T)e^{-i\omega_1 t}, & t < 0 \\
   e^{-i\omega_2 (t-T)} + \rho_r(T) e^{i\omega_2 (t-T)}, & t > T
 \end{cases},
\end{equation}
where
\begin{equation}
\rho_i(T)=\frac{1}{s(T)},~\rho_r(T)=\frac{r(T)}{s(T)}.
\end{equation}
The wave equation can then be expressed in first-order matrix form:
\begin{eqnarray}
\frac{d}{dt}\begin{pmatrix}
  u_1(t) \\
  u_2(t)
\end{pmatrix}=B(t)
               \begin{pmatrix}
  u_1(t) \\
  u_2(t)
\end{pmatrix},
\label{eq:uu}
\end{eqnarray}
with
\begin{align}
B(t)=\begin{pmatrix}
                 0 & \frac{1}{\mu(t)} \\
                 -\frac{c^2k_x^2}{\epsilon(t)} & 0
               \end{pmatrix},~u_2(t) = \mu(t)\frac{du_1}{dt}.
\end{align}
Boundary conditions at $t = 0$ and $t = T$ are obtained by matching $u_1$ and $u_2$:
\begin{align}
&u_1(0,T)=\rho_i(T),~u_2(0,T)=-i\omega_1\mu_1\rho_i(T),\nonumber\\
&u_1(T,T)=1+\rho_r(T),~u_2(T,T)=i\omega_2\mu_2[\rho_r(T)-1].
\label{eq:uu2}
\end{align}
These conditions combine into a linear system:
\begin{eqnarray}
G{\bf S}+H{\bf R}={\bf v},
\label{eq:uu3}
\end{eqnarray}
where
\begin{eqnarray}
&&{\bf S}=\begin{pmatrix}
  u_1(0,T) \\
  u_2(0,T)
\end{pmatrix},~{\bf R}=\begin{pmatrix}
  u_1(T,T) \\
  u_2(T,T)
\end{pmatrix},\nonumber\\
&&G=\begin{pmatrix}
  i\omega_1\mu_1 & 1 \\
  0 & 0
\end{pmatrix},
~H=\begin{pmatrix}
  0 & 0 \\
  i\omega_2\mu_2 & -1
\end{pmatrix},\nonumber\\
&&{\bf v}=\begin{pmatrix}
  0 \\
  2i\omega_2\mu_2
\end{pmatrix}.
\end{eqnarray}
The vectors ${\bf R}$ and ${\bf S}$ depend linearly on ${\bf v}$:
\begin{eqnarray}
{\bf R}=\tilde{R}{\bf v},~{\bf S}=\tilde{S}{\bf v},
\label{eq:uu4}
\end{eqnarray}
with matrix elements such as
\begin{align}
\tilde{R}_{12}=\frac{1+\rho_r}{2i\omega_2\mu_2},~\tilde{R}_{22}=\frac{\rho_r-1}{2},~\tilde{S}_{12}=\frac{\rho_i}{2i\omega_2\mu_2}.
\label{eq:uu5}
\end{align}
Applying the invariant imbedding framework, the differential equations for $\tilde{R}$ and $\tilde{S}$ are \cite{sk3}:
\begin{align}
&\frac{d}{d\tau}\tilde{R}(\tau)=B(\tau)\tilde{R}(\tau)-\tilde{R}(\tau)HB(\tau)\tilde{R}(\tau),\nonumber\\
&\frac{d}{d\tau}\tilde{S}(\tau)=-\tilde{S}(\tau)HB(\tau)\tilde{R}(\tau),
\label{eq:aiie1}
\end{align}
integrated from $\tau=0$ to $\tau=T$ with
\begin{align}
\tilde{R}(0)=\tilde{S}(0)=(G+H)^{-1}.
\end{align}
Finally, the evolution equations for $\rho_i$ and $\rho_r$ are
\begin{align}
&\frac{1}{ck_x}\frac{d\rho_i}{d\tau}=i\beta\rho_i+i\gamma\rho_i\rho_r,\nonumber\\
&\frac{1}{ck_x}\frac{d\rho_r}{d\tau}=2i\beta\rho_r+i\gamma(\rho_r^2+1),
\end{align}
where
\begin{align}
&\beta = \frac{1}{2n_2} \left[ \frac{\epsilon_2}{\epsilon(\tau)} + \frac{\mu_2}{\mu(\tau)} \right],~
\gamma = \frac{1}{2n_2} \left[ \frac{\epsilon_2}{\epsilon(\tau)} - \frac{\mu_2}{\mu(\tau)} \right],\nonumber\\
&n_2 = \sqrt{\epsilon_2\mu_2}.
\end{align}
Alternatively, the temporal transmission and reflection amplitudes $s(\tau)$ and $r(\tau)$ satisfy
\begin{align}
&\frac{1}{ck_x}\frac{ds}{d\tau} = -s^2\frac{1}{ck_x}\frac{d\rho_i}{d\tau} = -i\gamma r - i\beta s,\nonumber\\
&\frac{1}{ck_x}\frac{dr}{d\tau} = s\frac{1}{ck_x}\frac{d\rho_r}{d\tau} + \rho_r\frac{1}{ck_x}\frac{ds}{d\tau} = i\beta r + i\gamma s,
\label{eq:imbed1}
\end{align}
with initial conditions
\begin{align}
r(0) = \frac{1}{2} \left( 1 - \sqrt{\frac{\epsilon_2\mu_1}{\epsilon_1\mu_2}} \right), ~
s(0) = \frac{1}{2} \left( 1 + \sqrt{\frac{\epsilon_2\mu_1}{\epsilon_1\mu_2}}  \right).
\label{eq:imbed2}
\end{align}
Defining the temporal transmittance
$S$ and reflectance
$R$ as
\begin{align}
   S=\sqrt{\frac{\epsilon_1\mu_2}{\epsilon_2\mu_1}}\vert s\vert^2,~
   R=\sqrt{\frac{\epsilon_1\mu_2}{\epsilon_2\mu_1}}\vert r\vert^2,
 \end{align}
it follows directly from Eqs.~(\ref{eq:imbed1}) and (\ref{eq:imbed2}) that $S - R = 1$,
regardless of the temporal variations in $\epsilon$ and $\mu$. This relation embodies the momentum conservation law in time-varying electromagnetic systems.

\subsection*{Dirac waves}

We next extend the formulation to Dirac waves. Consider a $p$-band Dirac wave of unit amplitude, $\psi_1(t)=e^{-i\omega_1 t}$ for $t<0$, with spatial dependence $e^{ik_xx}$. When the vector potential ${\bf A}(t)$ varies within $0\le t\le T$, the wavefunction evolves as
\begin{equation}
\psi_1(t, T) =
\begin{cases}
e^{-i\omega_1 t}, & t < 0 \\
 s(T) e^{-i\omega_2 (t - T)}+r(T) e^{i\omega_2 (t - T)}, & t > T
\end{cases},
\end{equation}
where $\omega_i=v_F|q_i|$, and $q_i$ is defined in Eq.~(\ref{eq:qq}). The coefficients $s$ and $r$ correspond to intraband ($p\to p$) and interband ($p\to h$) scattering, interpreted as temporal transmission and reflection.

The invariant imbedding equations for $r$ and $s$ are derived analogously to the electromagnetic case and take the form \cite{kim2}
\begin{align}
\frac{1}{v_F k_x} \frac{dr}{d\tau} = i\tilde{\beta} r + \tilde{\gamma} s,~
\frac{1}{v_F k_x} \frac{ds}{d\tau} = -\tilde{\gamma} r - i\tilde{\beta} s,
\label{eq:imbed3}
\end{align}
where
\begin{align}
\tilde{\beta} = \frac{1 + \alpha\alpha_2 + (\alpha + \alpha_2)\cos\theta}{\tilde{\epsilon}_2}, ~
\tilde{\gamma} = \frac{(\alpha - \alpha_2)\sin\theta}{\tilde{\epsilon}_2},
\end{align}
and
\begin{align}
    \alpha=\frac{eA}{\hbar k},~\tilde{\epsilon}=\sqrt{1+\alpha^2+2\alpha\cos\theta}.
\end{align}
Here, $\theta$ denotes the angle between the wave vector and the vector potential, while $\tilde{\epsilon}_2$ and $\alpha_2$ represent the final values of $\epsilon$ and $\alpha$, respectively.
The initial conditions at $\tau = 0$ are
\begin{align}
r(0) = \frac{1}{2}\left(1 - \frac{f_2}{f_1}\right), ~
s(0) = \frac{1}{2}\left(1 + \frac{f_2}{f_1}\right),
\label{eq:imbed4}
\end{align}
where $f_i = q_i / |q_i|$ denotes the normalized complex wave vector in region $i$.
The reflectance $R$ and transmittance $S$, defined as the probability densities of the reflected and transmitted waves relative to the incident wave, are
\begin{align}
R = |r|^2, \quad S = |s|^2.
\end{align}
From Eqs.~(\ref{eq:imbed3}) and (\ref{eq:imbed4}), it can be shown that $S + R = 1$,
throughout the temporal evolution of $\alpha(t)$ \cite{kim2}.
This relation reflects the simultaneous conservation of momentum and charge in Dirac systems.
Accordingly, the fundamental distinction between Dirac and electromagnetic waves is expressed in their respective conservation laws: $S + R = 1$ in the Dirac case, and $S - R = 1$ in the electromagnetic case.

\end{document}